\documentclass[aps,prc,preprint,superscriptaddress,showpacs]{revtex4}

\usepackage[dvips]{graphicx}
\usepackage{amsmath,amssymb}
\usepackage{ulem}

\newcommand{\beq}{\begin{equation}}
\newcommand{\eeq}{\end{equation}}
\newcommand{\bea}{\begin{eqnarray}}
\newcommand{\eea}{\end{eqnarray}}

\def\la{\mathrel{\mathpalette\fun <}}
\def\ga{\mathrel{\mathpalette\fun >}}
\def\fun#1#2{\lower3.6pt\vbox{\baselineskip0pt\lineskip.9pt
  \ialign{$\mathsurround=0pt#1\hfil##\hfil$\crcr#2\crcr\sim\crcr}}}
\begin{document}

\title{The Brieva-Rook Localization of the Microscopic
Nucleon-Nucleus Potential}

\author{Kosho Minomo}
\email[]{minomo@phys.kyushu-u.ac.jp}
\affiliation{Department of Physics, Kyushu University, Fukuoka 812-8581, Japan}

\author{Kazuyuki Ogata}
\email[]{ogata@phys.kyushu-u.ac.jp}
\affiliation{Department of Physics, Kyushu University, Fukuoka 812-8581, Japan}

\author{Michio Kohno}
\email[]{kohno@kyu-dent.ac.jp}
\affiliation{Physics Division, Kyushu Dental College,
Kitakyushu 803-8580, Japan}

\author{Yoshifumi R. Shimizu}
\email[]{shimizu@phys.kyushu-u.ac.jp}
\affiliation{Department of Physics, Kyushu University, Fukuoka 812-8581, Japan}

\author{Masanobu Yahiro}
\email[]{yahiro@phys.kyushu-u.ac.jp}
\affiliation{Department of Physics, Kyushu University, Fukuoka 812-8581, Japan}

\date{\today}

\begin{abstract}
The nonlocality of the microscopic nucleon-nucleus
optical potential is commonly localized by the Brieva-Rook approximation.
The validity of the localization is tested for the proton+$^{90}$Zr
scattering at the incident energies from 65~MeV to
800~MeV. The localization is valid in the wide incident-energy range.
\end{abstract}

\maketitle

\section{Introduction}
\label{Introduction}
Microscopic understanding of nucleon-nucleus (NA) elastic scattering is
a long-standing fundamental subject in the nuclear reaction theory.
This is nothing but to solve the many-body scattering problem.
The many-body collision, however, can be approximately
described as a scattering
between two bodies interacting via a complex mean-field (optical) potential.
This optical potential is an important ingredient in theoretical
calculations of cross sections of elastic and inelastic scattering,
charge exchange and transfer reactions, and so on.
This means that
a good global optical model is
a powerful tool for predicting observables of NA scattering
for which no measurements exist, e.g., scattering of unstable nuclei from
proton target.

A reasonable way of getting the optical potential is to
calculate the NA folding potential with the nucleon-nucleon (NN)
$g$-matrix
interaction~\cite{Brieva-Rook,CEG,Rikus,Amos,Amos-2,Furumoto:2008zz}.
The interaction is first evaluated
in infinite nuclear matter and then folded into target (A) density by using
the local-density approximation.
The $g$-matrix interaction thus obtained is a
complex nonlocal potential depending on the incident energy ($E$)
of nucleon (N) and the nuclear-matter density ($\rho$).
This microscopic optical potential is successful in reproducing
data of NA elastic scattering~\cite{Amos} in the wide range of
$40 < E < 250$~MeV from light to heavy targets.
Above the pion production threshold, resonance and meson production effects
are evident. Recently, a bare NN interaction
was extrapolated to reproduce the NN scattering data to 2.5~GeV
by adding a complex potential phenomenologically~\cite{von-Geramb},
and the $g$-matrix interaction constructed
from the complex NN interaction was also successful in
reproducing the NA scattering at $40 < E < 800$~MeV~\cite{Amos-2}.

In many applications, use of a nonlocal NA potential is impractical.
For example, in $^{8}$B+A scattering the projectile easily breaks up
into $^7$Be and p. This projectile breakup processes are described by solving
the scattering problem of three-body system $^7$Be+p+A.
If all potentials are local in the system, this problem can be solved by
the method of continuum-discretized coupled
channels (CDCC)~\cite{CDCC-review,Ogata:2005xd}.
If the potential between p and A and/or the potential
between $^7$Be and A is nonlocal, this is not easy.
For such cases, use of an equivalent local potential is quite practical,
if it is accurate. Brieva and Rook (BR) proposed
an approximate form of the equivalent
local potential~\cite{Brieva-Rook}. This is commonly used
in many applications;
for example see Refs.~\cite{Furumoto:2008zz,Furumoto:2009zz}
and references therein. However, the validity of the approximate form is
not shown yet.

In this paper, we show the validity of the BR localization
over the wide range of $65 < E < 800$~MeV,
comparing the scattering solution of the non-local NA potential
with that of the BR-type local potential.
As a typical case, we consider the p+$^{90}$Zr scattering.
The BR localization is composed of three approximations.
We show that one of the three is redundant, and
test the remaining two separately.

In Sec.~\ref{formulation}, the method
of solving the Schr\"odinger equation with
the non-local NA potential and the way of
getting the ground-state wave function of target nucleus
are presented.
In Sec.~\ref{BR}, the BR localization is recapitulated.
In Sec.~\ref{Results}, the
validity of the BR localization and the related topics are argued.
Section~\ref{Summary} is devoted to summary.

\section{Formulation}
\label{formulation}

In the $g$-matrix approach~\cite{Brieva-Rook,CEG,Rikus,Amos,Furumoto:2008zz},
the microscopic NA optical potential is constructed
by folding the $g$-matrix interaction $g^{ST}$
with the ground-state density of target nucleus (A),
where $S$ ($T$) is the spin (isospin) of the N+N system.
In this procedure, the antisymmetrization between an incident nucleon and
target nucleons in A is taken care of
by using $g^{ST}$ which is properly antisymmetric
with respect to
the exchange of the colliding nucleons, since the prescription is shown to
be a good approximation~\cite{Takeda,Picklesimer}.
In the approximation, the folding potential is expressed
by the sum of
a local direct term $U^{\rm DR}(\mathbf{R})$ and a nonlocal exchange
term $U^{\rm EX}(\mathbf{R},\mathbf{r})$~\cite{Owen-Satchler}.
Hence, the elastic scattering
can be described by solving the Schr\"odinger equation
\begin{equation}
\left[  -\frac{\hbar^{2}}{2\mu}\nabla_{R}^{2}+U^{\rm DR}\left(\mathbf{R}\right)
+V_{\rm c} \left(  R\right)\delta^{\nu_1}_{-1/2}
-E\right]  \chi_{\mathbf{K},\nu_{1}}\left(  \mathbf{R}\right)  =
\int U^{\rm EX}(\mathbf{R},\mathbf{r})
\chi_{\mathbf{K},\nu_{1}}\left(  \mathbf{r}\right) d\mathbf{r}
\label{SchroeEqD}
\end{equation}
for the relative wave function
$\chi_{\mathbf{K},\nu_{1}}\left(  \mathbf{R}\right)$, where
$\mathbf{R}$ stands for
the coordinate of incident nucleon (N) from the center-of-mass of
A, $\mathbf{r}$ is the coordinate of nucleon in A
from the center-of-mass of A,
$V_{\rm c} \left(  R\right)$ is the Coulomb potential,
$\hbar \mathbf{K}$ ($E$) is an incident momentum (energy), and
$\nu_1=1/2$ for neutron scattering and $-1/2$ for proton one.
We assume that the target nucleus is much heavier than N.
The relativistic kinematics is taken by defining
the reduced mass $\mu$ as $\mu=\sqrt{m_{\rm N}^2+(p/c)^2}$ with
$p$ and $m_{\rm N}$ the momentum and rest mass of N.

In this paper, we consider only
the central part of the microscopic optical potential,
since it is a main component of the folding potential.
We also assume that the ground state of A is
described by a single determinant of
single-nucleon wave functions
$\varphi_{\nu_2;nljj_{z}}\left(  \mathbf{r},\xi\right)$,
each classified with the $z$-component $\nu_2$ of isospin,
the principal quantum number $n$, the angular momentum $l$, and the total
angular momentum $j$ and its $z$-component $j_z$,
where $\xi$ is the internal coordinate of
the spin wave function $\eta_{1/2}$ of nucleon.
In this case, $U^{\rm DR}(\mathbf{R})$ and $U^{\rm EX}(\mathbf{R},\mathbf{r})$
are given by~\cite{Brieva-Rook,CEG,Rikus,Amos}
\bea
U^{\rm DR}(\mathbf{R})&=&\sum_{\nu_{2},T_z}\int
\rho_{{\nu_{2}}}(\mathbf{r})
g_{T_{z}}^{\rm DR}(  s;\rho_{\nu_{2}} )
d\mathbf{r} ,
\label{UD} \\
{U}^{\rm EX}(\mathbf{R},\mathbf{r})
&=&
\sum_{\nu_{2},T_z}\rho_{{\nu_{2}}}(\mathbf{R},\mathbf{r})
g_{T_{z}}^{\rm EX}(  s;\rho_{\nu_{2}} )
\label{UE}
\eea
with
\bea
\rho_{{\nu_{2}}}(\mathbf{r})&=&
\sum_{nljj_{z}}\int\varphi_{\nu_{2};nljj_{z}}^{\ast}\left(\mathbf{r},\xi\right) \varphi_{\nu_{2};nljj_{z}}\left(
\mathbf{r},\xi\right) d\xi , \\
\rho_{{\nu_{2}}}(\mathbf{R},\mathbf{r})&=&
\sum_{nljj_{z}}\int\varphi_{\nu_{2};nljj_{z}}^{\ast}\left(\mathbf{r},\xi\right) \varphi_{\nu_{2};nljj_{z}}\left(
\mathbf{R},\xi\right) d\xi , \\
g_{T_{z}=\pm1}^{\rm DR}(s;\rho_{\nu_{2}})&=&
\frac{1}{4} \left\{  g^{01}\left(  s;\rho_{\nu_{2}}\right)
+3g^{11}\left(  s;\rho_{\nu_{2}}\right)  \right\}
\delta^{\nu_{1}+\nu_{2}}_{T_z} , \\
g_{T_{z}=0}^{\rm DR}(s;\rho_{\nu_{2}})&=&
\frac{1}{8}\left\{  g^{01}\left(  s;\rho_{\nu_{2}}\right)
+3g^{10}\left(  s;\rho_{\nu_{2}}\right)
+g^{00}\left(  s;\rho_{\nu_{2}}\right)
+3g^{11}\left(s;\rho_{\nu_{2}}\right)  \right\}
\delta^{\nu_{1}+\nu_{2}}_{T_z} , \\
g_{T_{z}=\pm1}^{\rm EX}\left(  s;\rho_{\nu_{2}}\right)  &=&
\frac{1}{4}\left\{  -g^{01}\left(s;\rho_{\nu_{2}}\right)
+3g^{11}\left(  s;\rho_{\nu_{2}}\right)  \right\}
\delta^{\nu_{1}+\nu_{2}}_{T_z} , \\
g_{T_{z}=0}^{\rm EX}\left(  s;\rho_{\nu_{2}}\right) &=&
\frac{1}{8}\left\{  -g^{01}\left(s;\rho_{\nu_{2}}\right)
-3g^{10}\left(  s;\rho_{\nu_{2}}\right)
+g^{00}\left(  s;\rho_{\nu_{2}}\right)
+3g^{11}\left(  s;\rho_{\nu_{2}}\right)  \right\}
\delta^{\nu_{1}+\nu_{2}}_{T_z},
\eea
where $\mathbf{s}=\mathbf{r}-\mathbf{R}$.
The $g$ matrix interaction $g^{ST}$ is a function of $E$
and the single-particle density $\rho_{\nu_{2}}(r_{g})$,
where $r_g=|\mathbf{r}_g|$ for the location $\mathbf{r}_g$
at which the effective
interaction works.
As for the $g$ matrix interaction, we take
a sophisticated version of the Melbourne interaction
\cite{Amos} that is constructed from the Bonn-B
NN potential \cite{BonnB} and
includes a modification due to the pion-production effect \cite{von-Geramb}.
Since the interaction has a finite range, $r_g$
can not be determined uniquely. Possible choices are
(i) $r_g=r$, (ii) $r_g=R$ and
(iii) $r_g=r_m \equiv |\mathbf{r}+\mathbf{R}|/2$.
This ambiguity is referred to as the $r_{g}$-ambiguity in this paper.
The $r_{g}$-ambiguity is small, as shown later in
Fig.~\ref{fig:elastic-Zr-rg}. We then take choice (i).
We will return to this point below.

Expanding $g_{T_{z}}^{\rm DR}$ into a series of multipoles,
\bea
g_{T_{z}}^{\rm DR}\left(  s;\rho\right)&=&\sum_{\lambda}4\pi
\frac{\left(  -\right)  ^{\lambda}}{\hat{\lambda}}g_{T_{z};\lambda}
^{\rm DR}\left(  r,R;\rho\right)  \left[  Y_{\lambda}\left(  \mathbf{\hat{R}
}\right)  \otimes Y_{\lambda}\left(  \mathbf{\hat{r}}\right)  \right]  _{00},
\label{mpe}
\eea
one can get a simple form of
\begin{align}
U^{\rm DR}(R)
=4\pi\sum_{\nu_{2},T_z}\int\rho_{\nu_{2}}\left(
r\right)  g_{T_{z};0}^{\rm DR}\left(  r,R;\rho_{\nu_{2}}\right)  r^{2}dr .
\end{align}
In this form, $U^{\rm DR}(R)$ is a function of $R=|\mathbf{R}|$.

The scattering wave function $\chi_{\mathbf{K},\nu_{1}}(\mathbf{r})$ is
expanded into partial waves $\chi_{K\nu_{1},L^{\prime}}(r)$:
\bea
\chi_{\mathbf{K},\nu_{1}}(\mathbf{r})
=\frac{4\pi}{Kr}\sum_{L^{\prime}M^{\prime}}
\chi_{K\nu_{1},L^{\prime}}(r)
i^{L^{\prime}}Y_{L^{\prime}M^{\prime}}^{\ast}
\left(  \mathbf{\hat{K}}\right)
Y_{L^{\prime}M^{\prime}}\left(  \mathbf{\hat{r}}\right) .
\label{PWEX-chi}
\eea
Inserting Eq.~(\ref{PWEX-chi}) into Eq.~(\ref{SchroeEqD}),
multiplying the equation by $Y^{*}_{LM}(\mathbf{\hat{R}})$ from the left and
integrating it over the solid angle $\mathbf{\hat{R}}$, one can get
an equation for $\chi_{K\nu_{1},L}$ as
\bea
& {\displaystyle
\left[  -\frac{\hbar^{2}}{2\mu}\frac{d^{2}}{dR^{2}}+\frac{\hbar^{2}}{2\mu
}\frac{L\left(  L+1\right)  }{R^{2}}+U^{\rm DR}\left(  R\right)
+V_{\rm c} \left(  R\right)\delta^{\nu_1}_{-1/2} -E\right]
\chi_{K\nu_{1},L}\left(  R\right) }
\notag \\
& {\displaystyle
=\sum_{\nu_{2},T_z;nlj\lambda}\frac
{\hat{\jmath}^{2}}{\hat{l}^{2}}\left(  L0\lambda0|l0\right)  ^{2}\int\phi
_{\nu_{2};nlj}^{\ast}\left(  r\right)  g_{T_{z};\lambda}^{\rm EX}\left(
r,R;\rho_{\nu_{2}}(r_g)\right)  \phi_{\nu_{2};nlj}\left(  R\right)
\chi_{K\nu_{1},L}\left(
r\right)  dr },
\label{Sch-exact}
\eea
where $g_{T_{z}}^{\rm EX}$ has been expanded into multipoles
$g_{T_{z};\lambda}^{\rm EX}$
just as
in Eq.~(\ref{mpe}), and $\phi_{\nu;nlj}(r)$ is the radial part
of the single-nucleon wave function
$\varphi_{\nu;nljj_{z}}(\mathbf{r},\xi)$.
In the derivation of Eq.~(\ref{Sch-exact}),
$\delta_{L^{\prime}L}$ came out in $L^{\prime}$ summation,
so that
$\chi_{K\nu_{1},L}(R)$
on the left hand side
has the same angular momentum $L$ as $\chi_{K\nu_{1},L}(r)$
on the right hand side. This is a consequence of the fact
that Eq.~(\ref{SchroeEqD}) is rotational invariant in the coordinate space.
In the present paper, $\chi_{K\nu_{1},L}(R)$ is obtained by solving
Eq.~(\ref{Sch-exact}) iteratively.
In Eq.~(\ref{Sch-exact}), each multipole
$g_{T_{z};\lambda}^{\rm EX}$ depends on $r_g$
through $\rho_{\nu_2}$, but it includes no information on an angle
between vectors $\mathbf{r}$ and $\mathbf{R}$.
Hence, we can not take choice (iii), which is commonly used
in the BR localization, in the form of Eq.~(\ref{Sch-exact}).
For this reason in addition to the reason that
the $r_g$-ambiguity itself is small,
we take choice (i) in the present study.

As for the ground state wave function of target nucleus,
it is desirable to be as realistic as possible,
and to be calculated theoretically
because we are planning to apply the formulation to unstable nuclei
where no experimental data are expected.
Therefore, we employ the Hartree-Fock (HF) calculation
with the finite-range Gogny force\cite{DeChGog}
as an effective interaction.
In particular, the D1S parameter-set\cite{BergGirGog} is adopted,
which is applied widely and successfully to many nuclear structure
problems (see, e.g., Ref.\cite{HilGir}).
The standard method to solve the HF equation with the Gogny force
is to expand the single-nucleon wave functions
in terms of the harmonic oscillator basis.
It is, however, not very accurate when the wave functions extend
far outside nucleus due to the weak-binding, which is characteristic
in unstable nuclei.
The Gaussian expansion method (GEM)~\cite{Kamimura,HiyKinKami}
is a powerful method
to treat such a problem of the spacially extended wave functions,
and it has been applied for solving the HF\cite{NakaSato} and
HFB (Hartree-Fock-Bogoliubov)\cite{Nakada} equations.
We have developed our own program to solve the HF and HFB equations
by the Gaussian expansion,
where the merit of the Gaussian form of the interaction is fully utilized.

In the present work, the target nucleus is $^{90}$Zr, which is a stable
nucleus and the method of the harmonic oscillator basis expansion works.
However, the use of the Gaussian basis expansion is still preferable
because the calculation of the matrix elements of the $g$-matrix
in Eq.~(\ref{Sch-exact}) can be done easily and accurately.
We have calculated the ground state of $^{90}$Zr by the HFB method.
It is found that the neutron pairing gap vanishes
because of the $N=50$ shell closure and
the proton pairing gap is also very small due to the subshell $Z=40$.
The energy gain by the pairing correlation is less than 100~keV,
and its effect on the density $\rho_{\nu}(r)$ is less than 1.5\%.
Therefore, we use the HF wave function neglecting the pairing correlation
in this work.
In more detail, the radial part of the single-nucleon wave functions
in each $(\nu;lj)$-channel are expanded by the Gaussian functions,
\beq
 \phi_{\nu;nlj}(r)=\sum_{i=1}^{n_g}
 C^{(\nu;lj)}_{n,i}\, e^{-(r/\lambda_i)^2},
\eeq
where we take $n_g=14$ and their ranges $\lambda_i$ ($i=1,..,n_g$)
are chosen to be from 1 to 5 fm by geometric progression
according to GEM\cite{Kamimura,HiyKinKami};
they are an almost optimal choice in the case of $n_g=14$.
The coefficients $C^{(\nu;lj)}_{n,i}$ are determined by
the HF variational equation.
The resultant binding energy of $^{90}$Zr is 785.995~MeV,
which is compared to the experimental value 783.894~MeV.
This result corresponds to the calculation employing 25 shells
($N_{\rm osc} \le 24$) in the harmonic oscillator basis.
We believe that the obtained wave function is realistic enough
to perform the test of the localization of optical potential.

\section{The Brieva-Rook localization}
\label{BR}

A local potential $U_{\rm loc}^{\rm EX}(\mathbf{R})$ trivially equivalent to
the nonlocal potential $U^{\rm EX}(\mathbf{R},\mathbf{r})$
is defined by
\bea
U_{\rm loc}^{\rm EX}(\mathbf{R})
\chi_{\mathbf{K},\nu_{1}}\left(  \mathbf{R}\right) =
\int U^{\rm EX}(\mathbf{R},\mathbf{r})
\chi_{\mathbf{K},\nu_{1}}\left(  \mathbf{r}\right) d\mathbf{r} .
\eea
Brieva and Rook derived an
approximate form $U_{\rm BR}^{\rm EX}(\mathbf{R})$ to the equivalent
local potential $U_{\rm loc}^{\rm EX}(\mathbf{R})$~\cite{Brieva-Rook}.
The derivation is composed of three approximations.
The first approximation, called the
local semi-classical approximation (LSCA) \cite{Luo-Kawai}, is
\begin{equation}
\chi_{\mathbf{K},\nu_{1}}\left(  \mathbf{r}\right)  =\chi_{\mathbf{K},\nu_{1}%
}\left(  \mathbf{R+s}\right)  \approx
\chi^{\rm LSC}_{\mathbf{K},\nu_{1}}\left(  \mathbf{r}\right)
\equiv
\chi_{\mathbf{K},\nu_{1}}\left(
\mathbf{R}\right)  e^{i \mathbf{K} \left(  \mathbf{R}\right)
\cdot\mathbf{s}},
\label{LSCA}
\end{equation}
where the local momentum $\hbar \mathbf{K} \left(  \mathbf{R}\right)$
is parallel to
the flux of the scattering wave at $\mathbf{R}$
and its magnitude is determined to satisfy
\beq
\hbar K(\mathbf{R})
=[2\mu(E-V_{\rm c}(R)\delta^{\nu_1}_{-1/2}
-U^{\rm DR}(\mathbf{R})-U_{\rm loc}^{\rm EX}(\mathbf{R}))]^{1/2},
\eeq
i.e., $\hbar \mathbf{K} \left(  \mathbf{R}\right)$ is evaluated
self-consistently.
LSCA has been successfully applied to studies on cross sections and
spin observables for multistep direct ($p,p'x$) and ($p,nx$)
processes as well as hyperon production cross sections~\cite{SCDW}.

LSCA yields a local form of
\bea
U_{\rm LSC}^{\rm EX}(\mathbf{R})
=\sum_{\nu_{2},T_z}\int
\rho_{{\nu_{2}}}(\mathbf{R},\mathbf{r})
 g_{T_{z}}^{\rm EX} (  s;\rho_{\nu_{2}}({r}_g) )
e^{i\mathbf{K}(\mathbf{R}) \cdot\mathbf{s}}d\mathbf{s} .
\label{U-EX-loc-1}
\eea
The local potential $U_{\rm LSC}^{\rm EX}(\mathbf{R})$ of
Eq.~(\ref{U-EX-loc-1})
is a function of the radial component $R$ and the angle $\theta$
between vectors $\mathbf{R}$ and $\mathbf{K}(\mathbf{R})$, as shown later.
Obviously, LSCA is getting better as $E$ increases.
Actually,
LSCA is good for $E \ga 65~{\rm MeV}$, as shown later in Sec.~\ref{Results}.

The second approximation, called the
local Fermi-gas approximation (LFGA)~\cite{Negele}, is an approximation
to the single-particle mixed density
$\rho_{{\nu_{2}}}(\mathbf{R},\mathbf{r})$:
\bea
\rho_{{\nu_{2}}}(\mathbf{R},\mathbf{r}) \approx
\rho_{{\nu_{2}}} \left(r_{\rm m}\right)
\frac{3}{(sk^{\rm F}_{{\nu_{2}}}(r_g))^3}
[\sin(sk^{\rm F}_{{\nu_{2}}}(r_g))-sk^{\rm F}_{{\nu_{2}}}(r_g)\cos(sk^{\rm F}_{{\nu_{2}}}(r_g))]
\equiv
\rho^{\rm LFG}_{{\nu_{2}}}(\mathbf{R},\mathbf{r}),
\label{LFGA}
\eea
where
$k^{\rm F}_{{\nu_{2}}}(r_g)$ is related to $\rho_{{\nu_{2}}}(r_g)$ as
\beq
\rho_{{\nu_{2}}}=\frac{(k^{\rm F}_{{\nu_{2}}})^3}{3\pi^2}.
\label{Fmom}
\eeq
LFGA is known to be a good approximation for small values of
$s$~\cite{Negele}.
The third approximation is expressed by
\bea
e^{i \mathbf{K}(\mathbf{R}) \cdot \mathbf{s}} \approx j_0(K(\mathbf{R})s),
\label{SSA}
\eea
where $j_{X}$ is the spherical Bessel function.
This approximation is good when
$j_0(K(\mathbf{R})s) \ll j_{X}(K(\mathbf{R})s) $ for $X \ge 1$.
This condition is well satisfied when $K(\mathbf{R})s \la 1$.
In Eq.~(\ref{U-EX-loc-1}), the range of the integrand
is about 0.5~fm because of the
presence of the short-ranged interaction $g_{T_{z}}^{\rm EX}$.
In the surface region of A that is important for forward NA scattering,
$K(\mathbf{R})$ approximately equals the asymptotic wave number $K$.
Hence, Eq.~(\ref{SSA}) is good at least
for $K \la 2~{\rm fm}^{-1}$ ($E \la 80~{\rm MeV}$).
Eventually, the BR-type equivalent local potential
$U_{\rm BR}^{\rm EX}(\mathbf{R})$ is obtained by
\bea
U_{\rm BR}^{\rm EX}(R)=
\sum_{\nu_{2},T_z}
\int
\rho^{\rm LFG}_{{\nu_{2}}}(\mathbf{R},\mathbf{r})
g_{T_{z}}^{\rm EX} (  s;\rho_{\nu_{2}}(r_g) )
j_0(K(R)s) d\mathbf{s} .
\label{U-EX-loc-2}
\eea

In the above derivation, LSCA is good for $E \ga 65~{\rm MeV}$, while
Eq.~(\ref{SSA}) is applicable for $E \la 80~{\rm MeV}$. However,
as shown in Sec.~\ref{sec:elastic}, the BR-type local potential
$U_{\rm BR}(R)\equiv U^{\rm DR}(R)+U_{\rm BR}^{\rm EX}(R)$ gives almost
the same elastic-scattering cross section as that obtained by
the exact calculation for $65 \le E \le 800$~MeV.
This means that the above derivation is not sufficient.
Actually, as shown in Sec.~\ref{sec:BR-localization},
Eq.~\eqref{U-EX-loc-2} is derivable from Eq.~\eqref{U-EX-loc-1} without
the approximation \eqref{SSA}, that is, the approximation is
redundant.
Although, in the original work~\cite{Brieva-Rook} of Brieva and Rook,
the local momentum
$\hbar \mathbf{K} \left(  \mathbf{R}\right)$
is assumed to be parallel to the asymptotic momentum $\hbar \mathbf{K}$,
it is not necessary because Eq.~\eqref{U-EX-loc-2} does not depend on
the direction of $\mathbf{K} \left(  \mathbf{R}\right)$.
Since LSCA itself is more accurate
for $\hbar \mathbf{K} \left(  \mathbf{R}\right)$ parallel to
the flux of the scattering wave at $\mathbf{R}$,
we should think that the direction is also taken
in the BR localization.

\section{Results}
\label{Results}

\subsection{Proton elastic-scattering from $^{90}$Zr}
\label{sec:elastic}

\begin{figure}[htb]
\begin{center}
 \includegraphics[clip,width=0.95\textwidth]{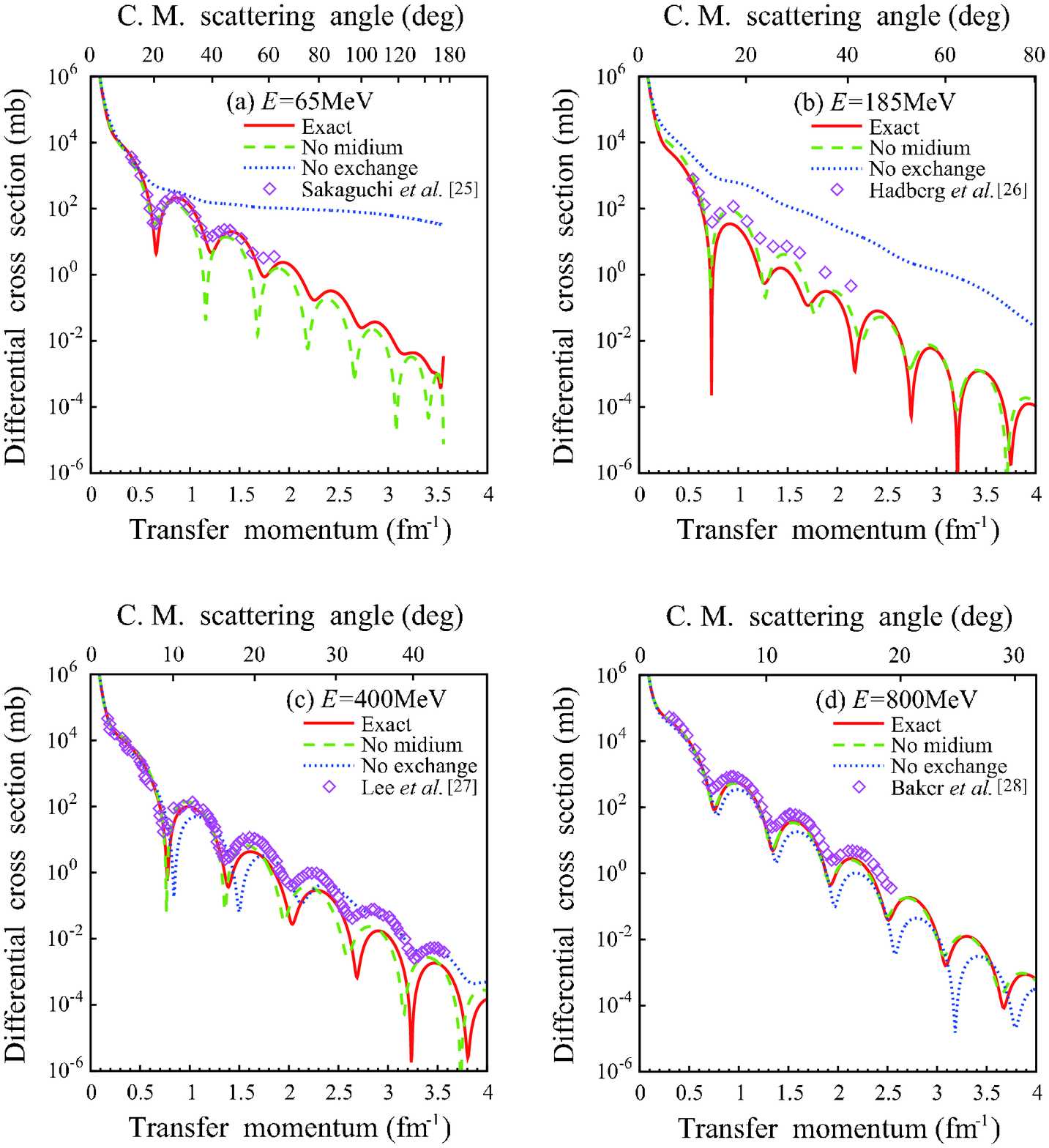}
 \caption{(color online) The differential cross sections of the
 proton scattering from $^{90}$Zr at
(a) $E=65$~MeV, (b) 185~MeV, (c) 400~MeV and (d) 800~MeV.
The solid curves represent the results of the exact calculation,
while the dashed (dotted) curves corresponding to the calculation
without the medium (exchange) effect.
The horizontal lower (upper) scale shows
the transferred wave number $q$ (the scattering angle ${\theta_{\rm cm}}$).
Experimental data are taken from Refs. \cite{Sakaguchi,Hadberg,Lee,Baker}.
 }
 \label{fig:elastic-Zr-DR}
\end{center}
\end{figure}

Figure \ref{fig:elastic-Zr-DR} presents the differential cross sections
of the  proton elastic scattering from $^{90}$Zr at (a) $E=65$~MeV,
(b) 185~MeV, (c) 400~MeV and (d) 800~MeV.
For each panel, the horizontal lower (upper) scale represents
the transferred wave number $q$
(the scattering angle ${\cal \theta_{\rm cm}}$).
The solid curves represent results of the exact calculation in which
Eq.~(\ref{SchroeEqD}) is solved numerically.
In the dashed curves, the medium effect is switched off
from the exact calculation by
taking $g^{ST}(s;\rho_{\nu_2}=0)$ in Eqs.~(\ref{UD}) and (\ref{UE}).
In the dotted curves, the exchange effect is neglected from
the exact calculation by setting $g_{T_z;\lambda}^{\rm EX}=0$
in Eq.~(\ref{SchroeEqD}).
Thus, the exchange effect is large at least up to $E=800$~MeV, and
the medium effect is significant up to $E=400$~MeV.

%%------------------------------------------------------------------%%
%% Figure
%%------------------------------------------------------------------%%
%%------------------------------------------------------------------%%

\begin{figure}[htb]
\begin{center}
 \includegraphics[clip,width=0.95\textwidth]{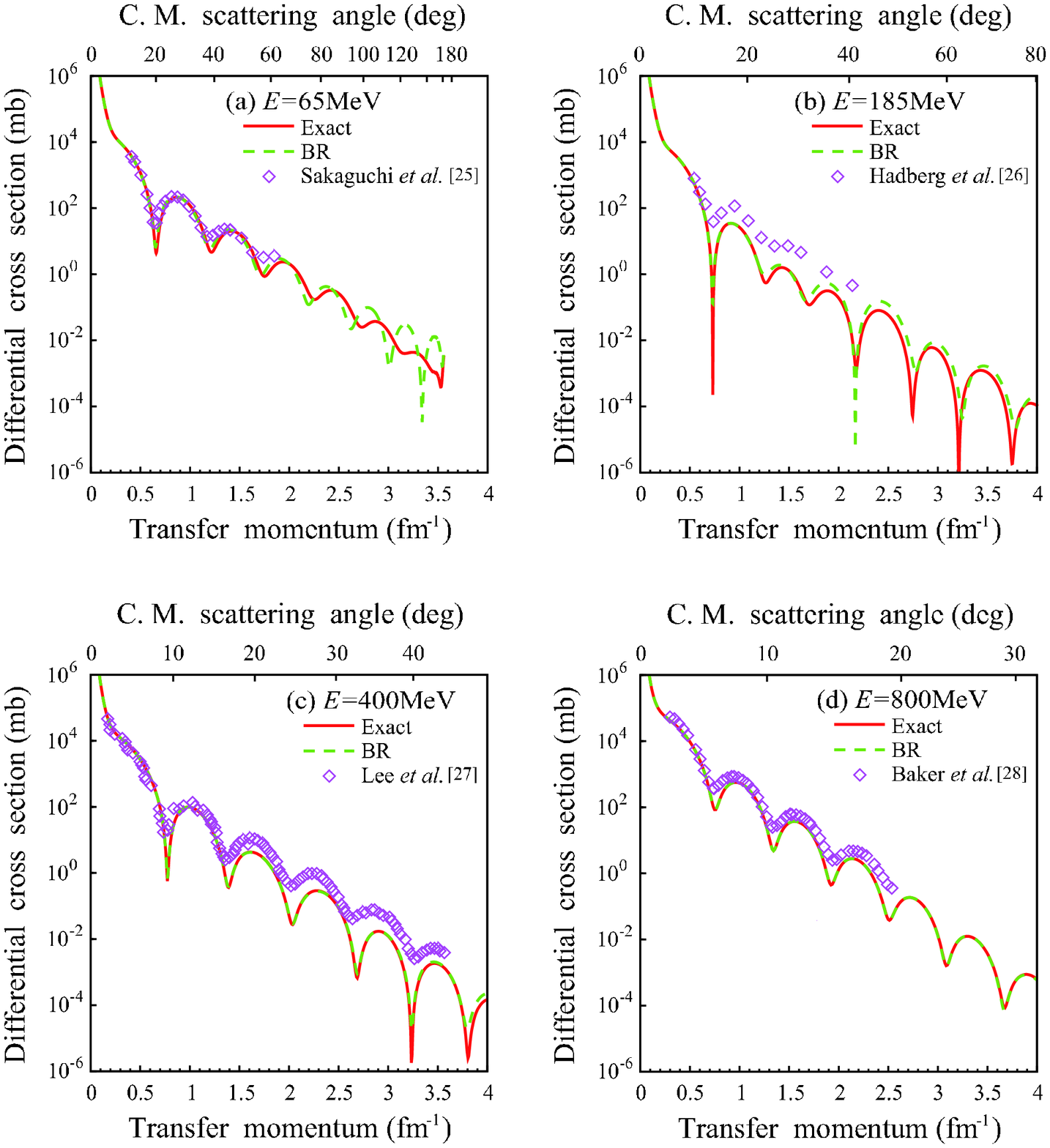}
 \caption{(color online) Same as in Fig.~\ref{fig:elastic-Zr-DR}
 except that the dashed lines show the results of the BR folding
 model, i.e., with $U_{\rm BR}(R)$.
 }
 \label{fig:elastic-Zr}
\end{center}
\end{figure}

Now, the validity of the BR localization is tested.
Figure~\ref{fig:elastic-Zr} presents the same quantities as in
Fig.~\ref{fig:elastic-Zr-DR}, but with different calculations.
The solid curves represent results of the exact calculation of
Eq.~(\ref{SchroeEqD}), while
the dashed curves do results of the BR-type local potential $U_{\rm BR}(R)$,
that is, the Schr\"odinger equation with
$U_{\rm BR}(R)\equiv U^{\rm DR}(R)+U_{\rm BR}^{\rm EX}(R)$
is solved numerically.
Seeing the difference between the two types of lines
around $q=3.5~{\rm fm}^{-1}$,
one can find that the error of the BR localization
is getting small as $E$ increases.
For $E=65$~MeV, the error is small at
$q \la 1.7~{\rm fm}^{-1}$ where the data are available, although
it is sizable at large $q$ around $3.5~{\rm fm}^{-1}$.
Thus, the BR localization is good for 65~MeV $\le E \le$ 800~MeV.
We discuss this point in Sec.~\ref{sec:BR-localization} in detail.

%%------------------------------------------------------------------%%
%% Figure
%%------------------------------------------------------------------%%
%%------------------------------------------------------------------%%

\begin{figure}[htb]
\begin{center}
 \includegraphics[clip,width=0.95\textwidth]{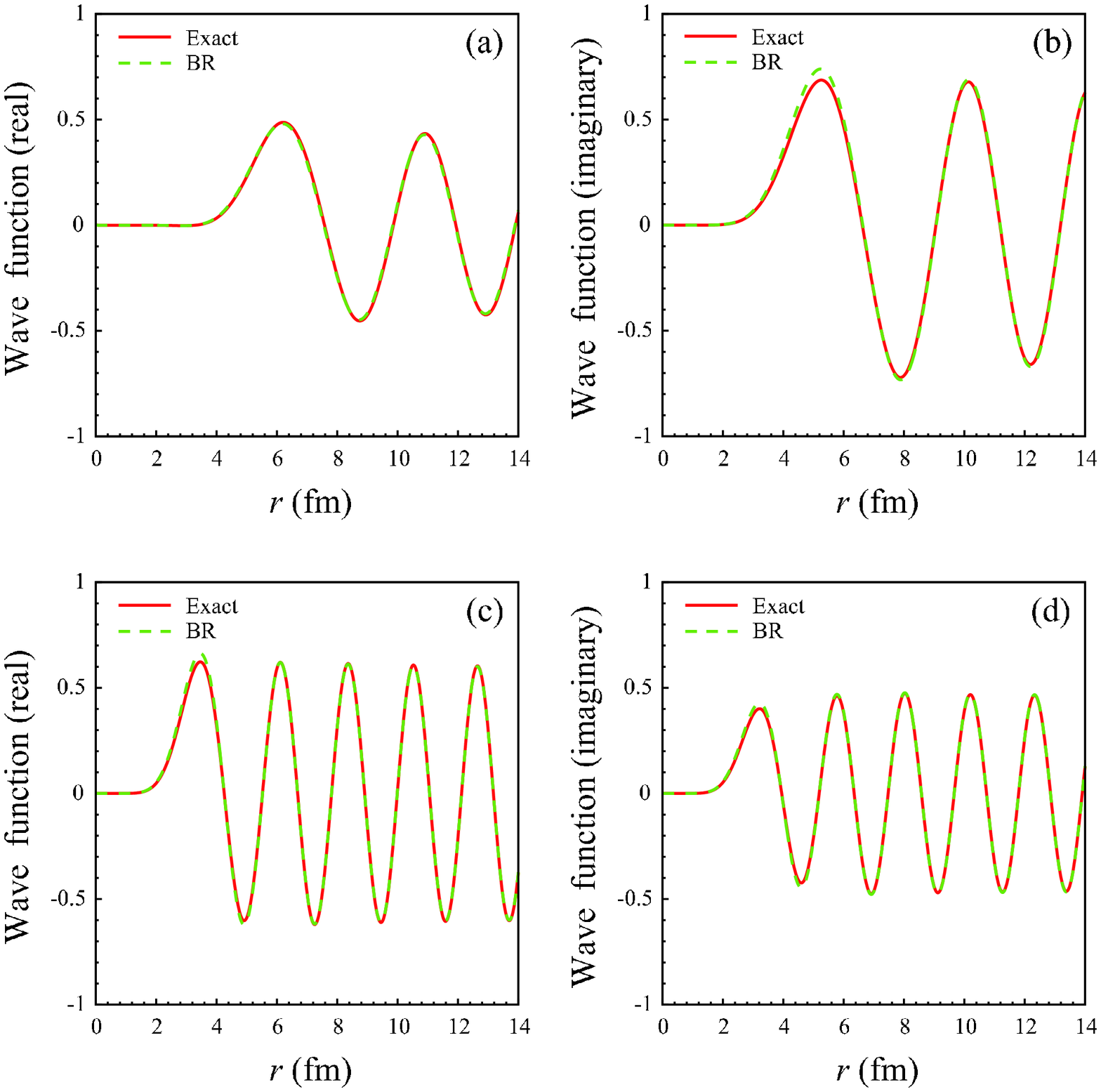}
 \caption{(color online)
 The exact and approximate wave functions for the p+$^{90}$Zr scattering
 in the case of $L=8$.
 The left and right panels represent the real and imaginary parts of
the wave functions, respectively. The upper (lower) panels correspond
to $E=65$~MeV (185~MeV).
 }
 \label{fig:WF}
\end{center}
\end{figure}

In Fig.~\ref{fig:WF}, the exact wave function (solid curve) of
the Schr\"odinger equation Eq.~(\ref{SchroeEqD})
is compared with the approximate wave function calculated
with $U_{\rm BR}(R)$
for the case of the p+$^{90}$Zr scattering at $L=8$,
where the elastic partial cross section
$\sigma_L^{\rm el}=\pi(2L+1)|1-S_L|^2/K^2$ calculated with
the elastic $S$-matrix element $S_L$ becomes maximum.
For both $E=65$~MeV and $185$~MeV,
the approximation wave functions are very close to the exact ones.

%%------------------------------------------------------------------%%
%% Figure
%%------------------------------------------------------------------%%
%%------------------------------------------------------------------%%

\subsection{Validity of the BR localization}
\label{sec:BR-localization}

First, we show the numerical test of LSCA, together with 
a simplified version of LSCA in which 
the absolute value of the local
momentum, $\hbar {K} \left(  \mathbf{R}\right)$, 
is replaced by that of the asymptotic momentum $\hbar K$. 
This version is referred to as LSCA-A in this paper. 
LSCA-A is, if justified, very useful since LSCA-A makes
it much simpler the numerical task
to obtain $U_{\rm BR}^{\rm EX}(R)$. 

For this purpose, we consider the potential scattering and take
$U_{\rm BR}(R)=U^{\rm DR}(R)+U_{\rm BR}^{\rm EX}(R)$ as the potential.
The exact wave function
$\chi_{\mathbf{K},\nu_{1}}\left(  \mathbf{r}\right)$
of the potential scattering is compared with the approximate wave functions
$\chi^{\rm LSA}_{\mathbf{K},\nu_{1}}\left(  \mathbf{r}\right)$ and
$\chi^{\rm LSC\mbox{-}A}_{\mathbf{K},\nu_{1}}\left(  \mathbf{r}\right)$
based on LSCA and LSCA-A, respectively.
The wave functions are invariant under the rotation around the $z$ axis,
and hence the azimuthal angle $\varphi$ of vector $\mathbf{R}$ can be
set to zero.
As an example, vector $\mathbf{R}$ is fixed to
$(R,\theta,\varphi)=(5~{\rm fm},\pi/3,0)$,
and vector $\mathbf{s}$ in Eq.~(\ref{LSCA}) is varied in
a direction either parallel or perpendicular to $\mathbf{R}$.
For convenience,
a variation of vector $\mathbf{s}$
in the direction parallel (perpendicular) to $\mathbf{R}$
is denoted by $s_r$ ($s_{\theta}$);
precisely,
$\mathbf{s_r}=(\mathbf{s} \cdot \mathbf{n})\mathbf{n}$
with $\mathbf{n}=\mathbf{R}/R$
and
$\mathbf{s}_{\theta}=\mathbf{s} - s_r \mathbf{n}$.
Note that LSCA is expected to work well if the potential around
$\mathbf{R}$ varies slowly within the wave length
of $\chi_{\mathbf{K},\nu_{1}}\left(  \mathbf{r}\right)$.
Thus, the choice of $R=5$ fm can severely test the validity of LSCA.

\begin{figure}[htb]
\begin{center}
 \includegraphics[clip,width=0.95\textwidth]{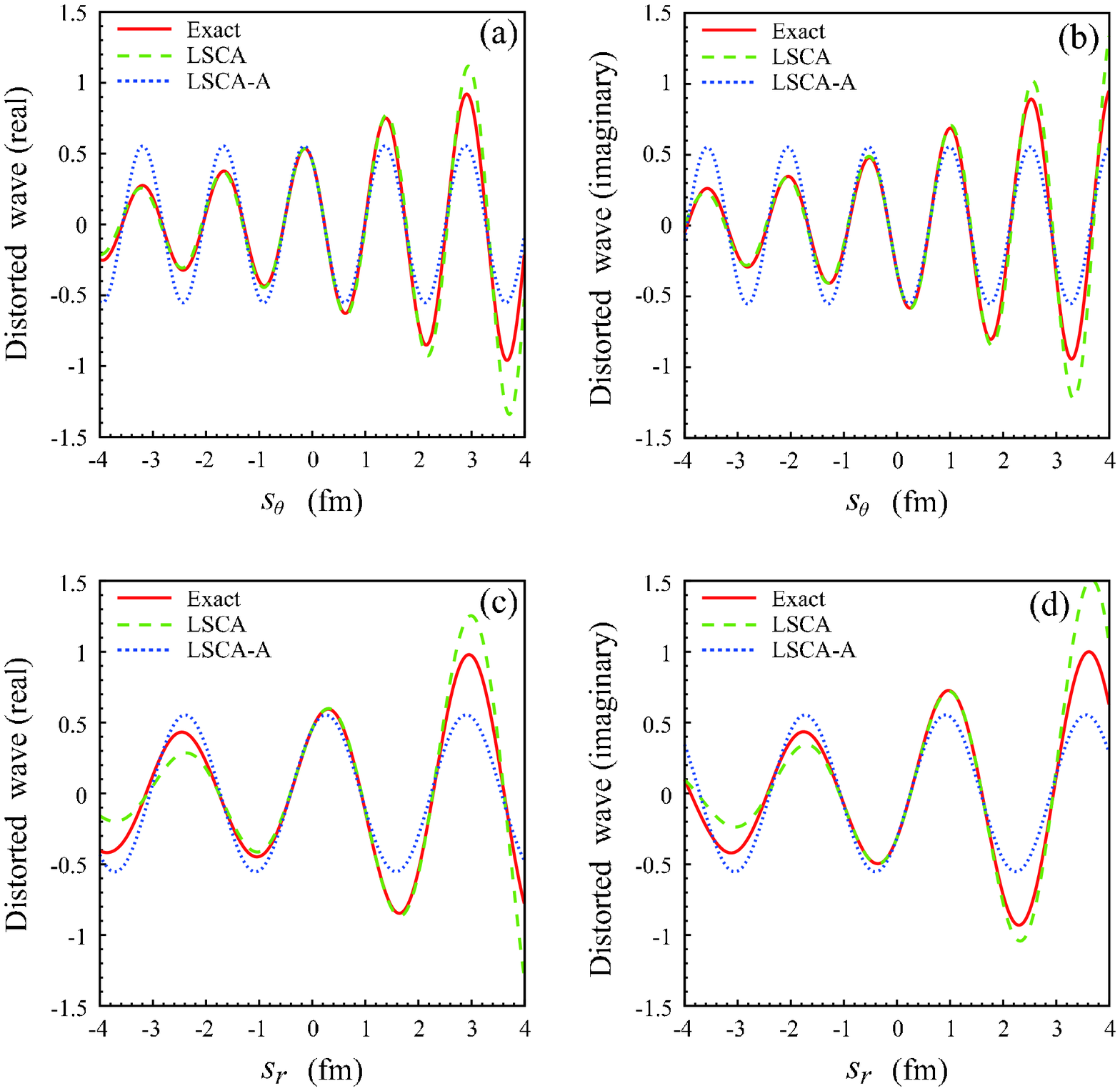}
 \caption{The exact and approximate wave functions of
 the p+$^{90}$Zr elastic scattering at $E=400$~MeV.
 Vector $\mathbf{R}$ is fixed at
 $\mathbf{R}=(R,\theta)=(5~{\rm fm}, \pi/3)$.
 In panels (a) and (b) $s_{\theta}$ is varied, while
 in panels (c) and (d) $s_r$ is varied.
 The left (right) panels represent the real (imaginary) parts
 of the wave functions. 
 In LSCA, the direction of vector 
 $\mathbf{K}(\mathbf{R})$  
 is assumed to be pararell to 
 the flux of the scattering wave at $\mathbf{R}$. The same assumption 
 is made also for the direction of $\mathbf{K}$ in LSCA-A. 
    }
 \label{fig:LSCA-E400}
\end{center}
\end{figure}

\begin{figure}[htb]
\begin{center}
 \includegraphics[clip,width=0.95\textwidth]{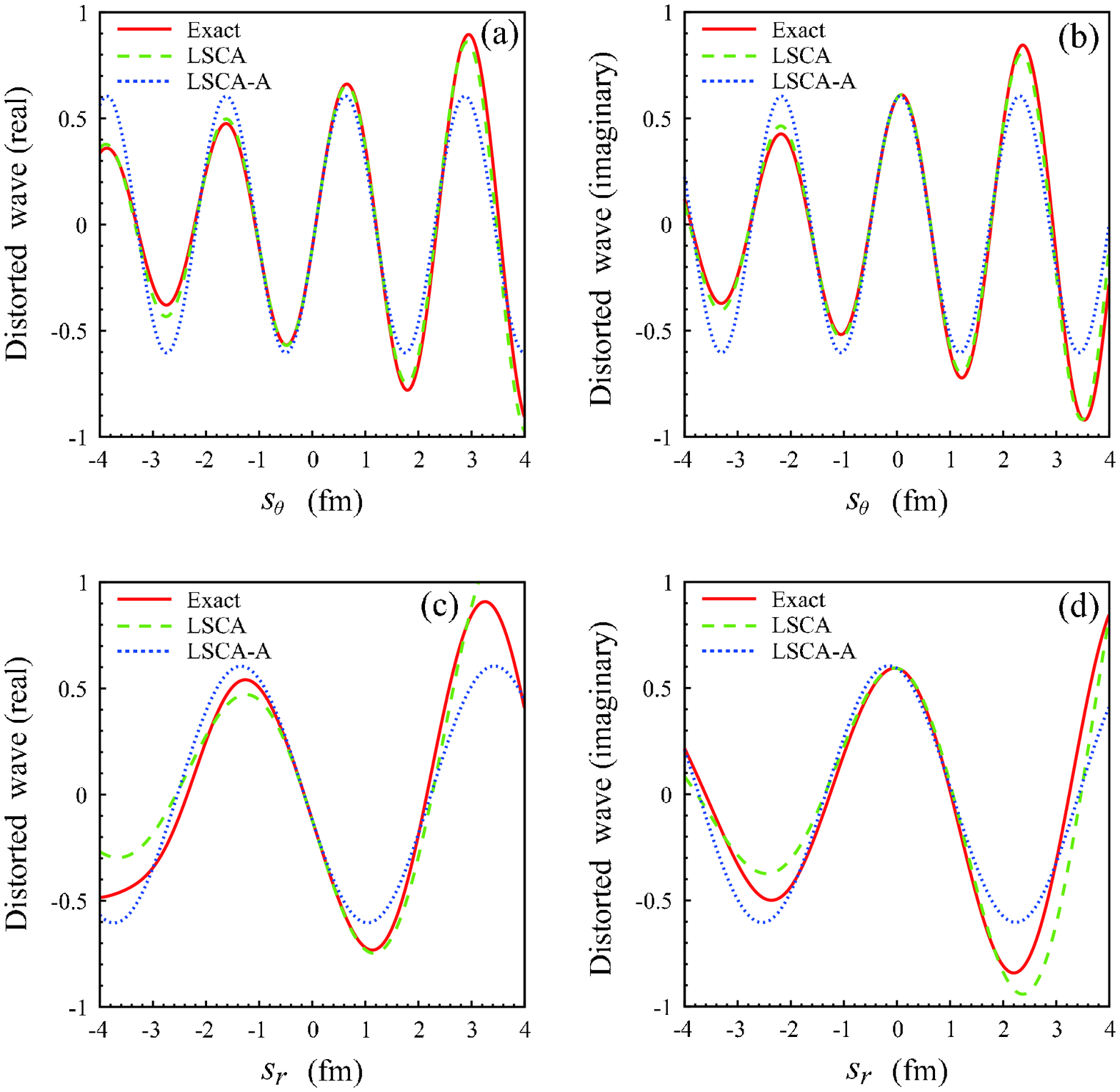}
 \caption{Same as in Fig.~\ref{fig:LSCA-E400} but at $E=185$~MeV.
   }
 \label{fig:LSCA-E185}
\end{center}
\end{figure}

\begin{figure}[htb]
\begin{center}
 \includegraphics[clip,width=0.95\textwidth]{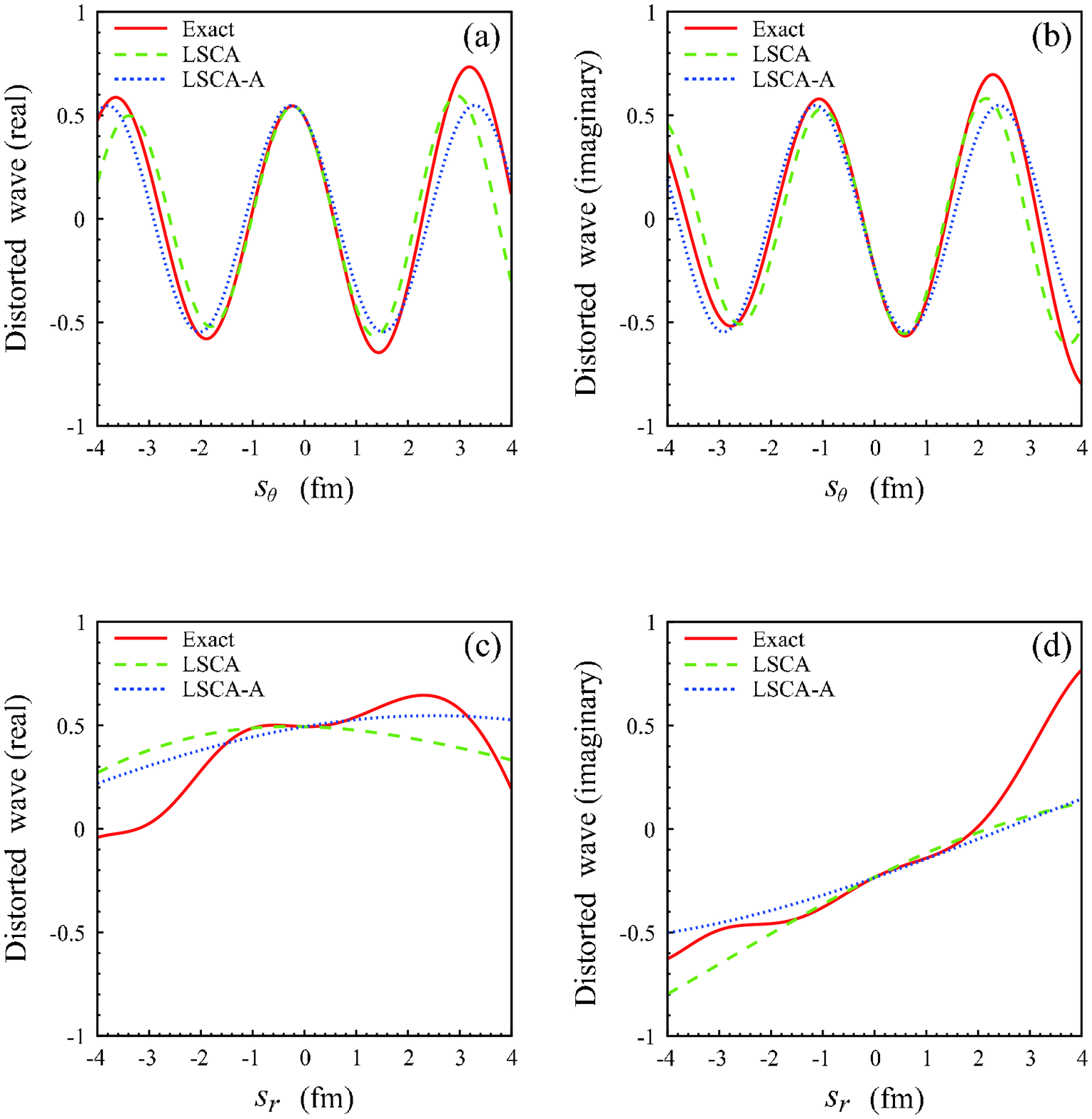}
 \caption{Same as in Fig.~\ref{fig:LSCA-E400} but at $E=65$~MeV.
   }
 \label{fig:LSCA-E65}
\end{center}
\end{figure}

\begin{figure}[htb]
\begin{center}
 \includegraphics[clip,width=0.95\textwidth]{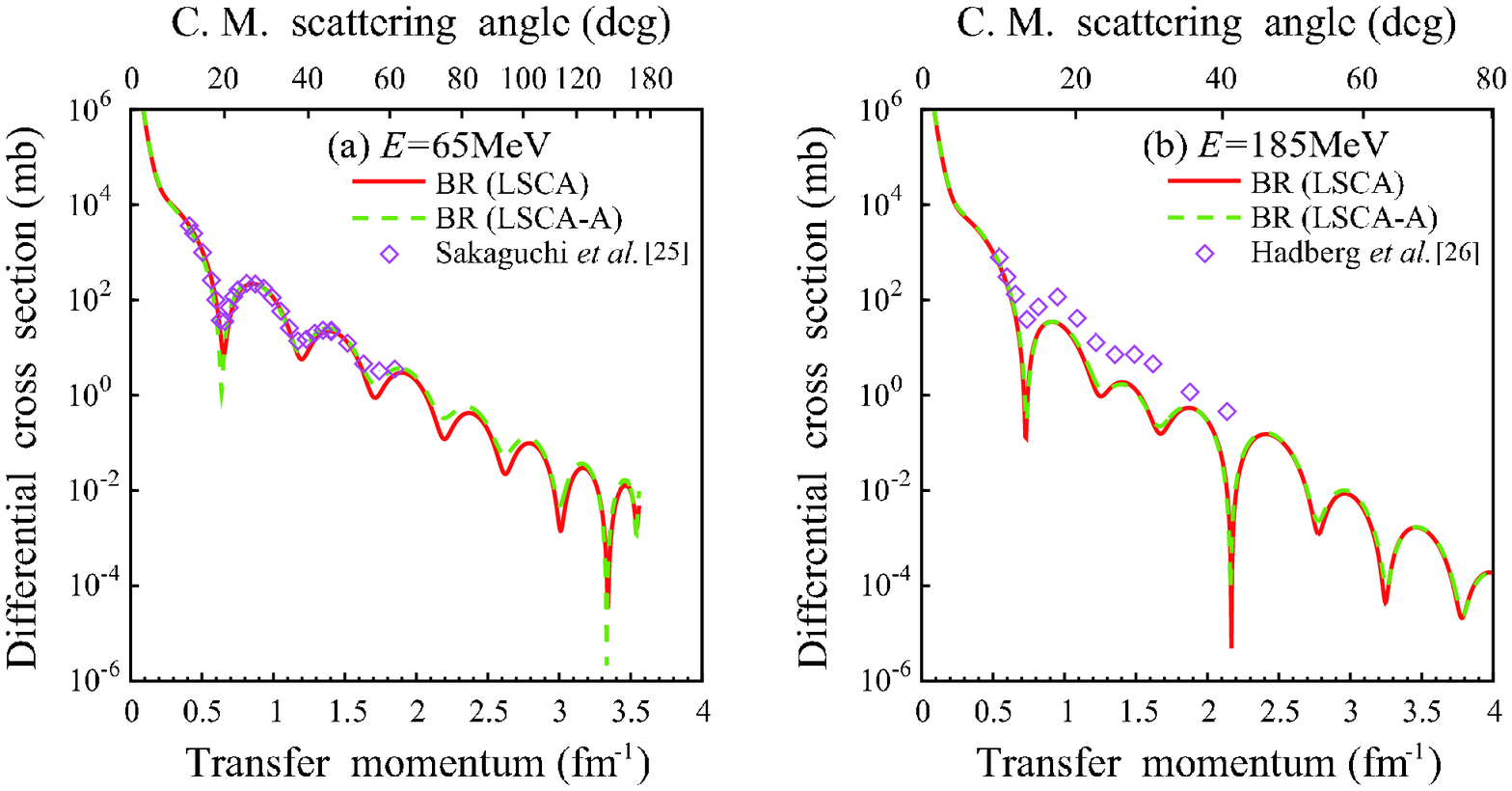}
 \caption{(color online)
 The validity of LSCA-A. Panels (a) and (b) represent
 the differential cross sections of the proton scattering from $^{90}$Zr at
$E=65$~MeV and 185~MeV, respectively. 
The dashed curves denote the results of the BR local potential in which
LSCA-A is taken instead of LSCA,
while the solid curves show the results of the ordinary BR local potential.
 }
 \label{fig:elastic-Zr-LSCA-A}
\end{center}
\end{figure}

Figure \ref{fig:LSCA-E400} represents the exact and approximate
wave functions of the p+$^{90}$Zr elastic scattering at $E=400$~MeV.
In panels (a) and (b) where $s_{\theta}$ is varied with $s_r$  fixed to 0,
the wave functions
$\chi^{\rm LSC}_{\mathbf{K},\nu_{1}}\left(  \mathbf{r}\right)$
(dashed curves)
and
$\chi^{\rm LSC\mbox{-}A}_{\mathbf{K},\nu_{1}}\left(  \mathbf{r}\right)$
(dotted curves)
agree with $\chi_{\mathbf{K},\nu_{1}}\left(  \mathbf{r}\right)$
(solid curves)
within the range of the $g$-matrix interaction, i.e. at $s \la 1.5~$fm.
This is the case also for panels (c) and (d) where $s_r$ is varied
with $s_{\theta}$  fixed to 0.
Further, $\chi^{\rm LSC}_{\mathbf{K},\nu_{1}}\left(  \mathbf{r}\right)$
is identical to 
$\chi^{\rm LSC\mbox{-}A}_{\mathbf{K},\nu_{1}}\left(  \mathbf{r}\right)$
at $s \la 1.5~$fm. Thus, LSCA and LSCA-A are accurate for high $E$.
At slightly lower energy of 185 MeV, as shown in
Fig.~\ref{fig:LSCA-E185}, the accuracy of LSCA and LSCA-A is
almost the same as at 400 MeV.
Figure \ref{fig:LSCA-E65} shows the results at 65 MeV.
One sees that LSCA and LSCA-A still work well, although its accuracy
becomes slightly worse than for $E \ge 185$ MeV. 
Thus, we conclude
that LSCA and LSCA-A are applicable for $E \ga 65$~MeV. 
Actually, as shown in Fig.~\ref{fig:elastic-Zr-LSCA-A}, 
the elastic-scattering cross section for
p$+^{90}$Zr at 65 and 185 MeV calculated with LSCA-A in the 
BR localization agrees well with the result of the standard BR
calculation with LSCA; particularly at 185 MeV, 
the difference between the two is within the thickness of lines.

%%------------------------------------------------------------------%%
%% Figure
%%------------------------------------------------------------------%%
%%------------------------------------------------------------------%%

%%------------------------------------------------------------------%%
%% Figure
%%------------------------------------------------------------------%%
%%------------------------------------------------------------------%%

%%------------------------------------------------------------------%%
%% Figure
%%------------------------------------------------------------------%%
%%------------------------------------------------------------------%%

%%------------------------------------------------------------------%%
%% Figure
%%------------------------------------------------------------------%%
%%------------------------------------------------------------------%%

Next, we consider the remaining two approximations, i.e.,
LFGA and Eq.~(\ref{SSA}). One can find, however, the latter is redundant,
if an extended LFGA below is justified. We consider the
following approximation for the mixed density:
\bea
\rho_{{\nu_{2}}}(\mathbf{R},\mathbf{r}) \approx
\rho_{{\nu_{2}}} \left(R \right)
\frac{3}{(sk^{\rm F}_{{\nu_{2}}}(R))^3}
[\sin(sk^{\rm F}_{{\nu_{2}}}(R))-sk^{\rm F}_{{\nu_{2}}}(R)\cos(sk^{\rm F}_{{\nu_{2}}}(R))]
\equiv
\rho^{{\rm LFG\mbox{-}}R}_{{\nu_{2}}}(\mathbf{R},\mathbf{r}).
\label{LFGA-R}
\eea
This approximation is referred to as LFGA-$R$.
Note that $\rho^{{\rm LFG\mbox{-}}R}_{{\nu_{2}}}(\mathbf{R},\mathbf{r})$
is nothing but $\rho^{\rm LFG}_{{\nu_{2}}}(\mathbf{R},\mathbf{r})$
of Eq.~(\ref{LFGA}), with $r_{\rm m}$ and $r_{g}$ replaced by $R$.
Obviously, as long as LFGA is valid, LFGA-$R$ is also good for $R$
larger than the range of the NN interaction $s \approx 0.5$~fm.
Inserting Eq.~(\ref{LFGA-R}) into
Eq.~(\ref{U-EX-loc-1}) leads to Eq.~(\ref{U-EX-loc-2}) without
Eq.~(\ref{SSA}),
since $\rho_{{\nu_{2}}}(\mathbf{R},\mathbf{r})$
is a function of $R$ and $s$ in LFGA-$R$.
Thus, Eq.~(\ref{SSA}) is not necessary, when LFGA-$R$ is taken.

The validity of LFGA-$R$ is evaluated by comparing the results
of the elastic-scattering cross sections calculated with LFGA and LFGA-$R$.
It is found that in the energy region of 65 MeV $\le E \le$ 800 MeV,
the two calculations show a perfect agreement.
Thus, we conclude that Eq.~(\ref{SSA}),
which has imposed an upper limit of $E$ where the BR localization is
accurate,
is actually redundant because of the good accuracy of LFGA-$R$.
This comes from the fact that, as mentioned above,
LFGA itself is a very good approximation to the mixed density.
In fact, it turns out that the p$+^{90}$Zr elastic-scattering
cross section calculated with LFGA agrees very well with that obtained with
explicitly using the mixed density.
Another point to be mentioned here is that the
simplest formula, Eq.~(\ref{Fmom}), is used for evaluating
$k^{\rm F}_{{\nu_{2}}}$ from the one-body density.
Our finding clearly shows that Eq.~(\ref{Fmom}) is enough to
study the elastic-scattering cross sections; the higher-order
corrections to $k^{\rm F}_{{\nu_{2}}}$ \cite{Negele,Campi} are not
necessary for this purpose.

\begin{figure}[htb]
\begin{center}
 \includegraphics[clip,width=0.95\textwidth]{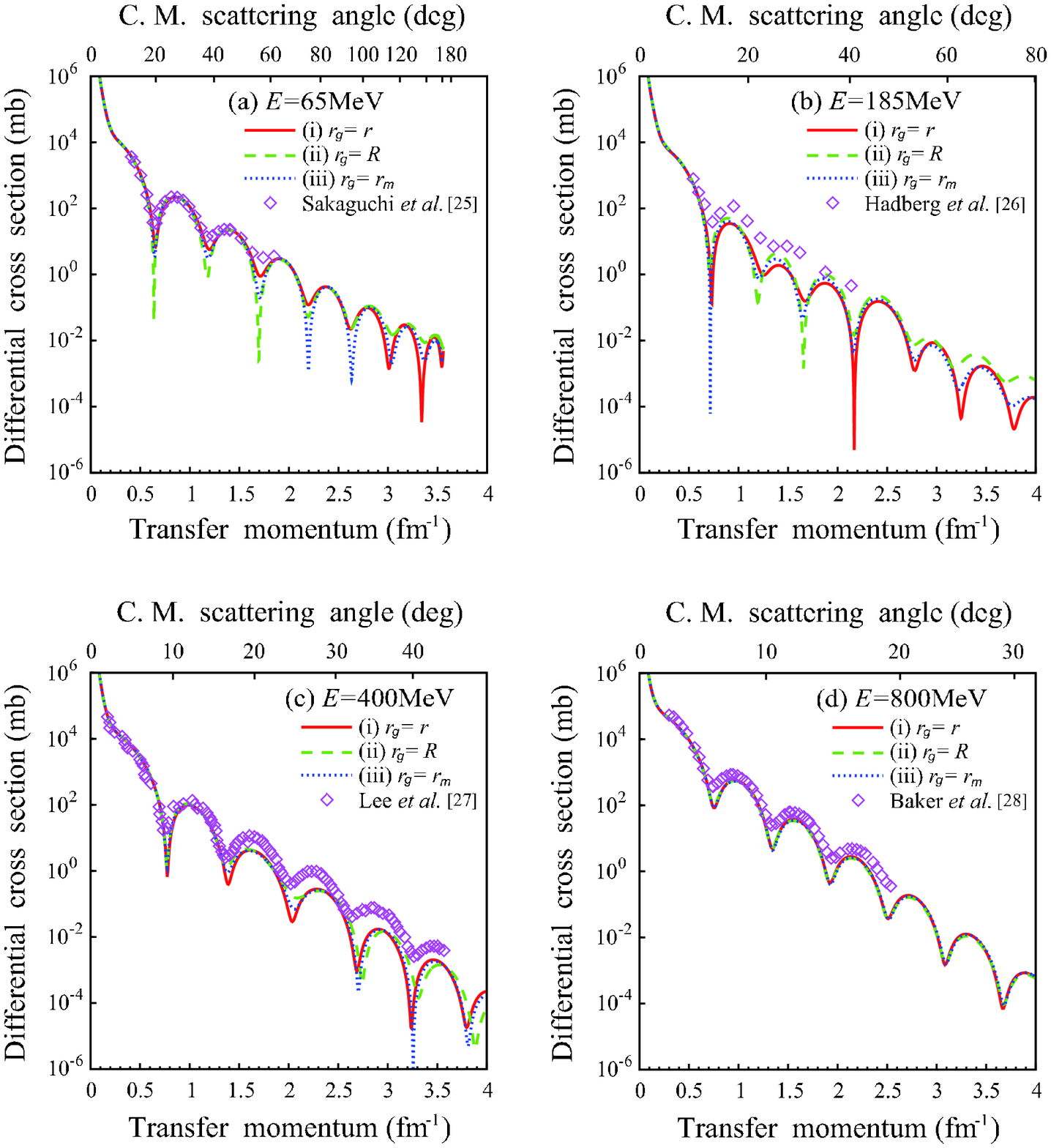}
 \caption{(color online)
 The ${r}_g$-ambiguity of the BR local potential for
the elastic-scattering cross section
from $^{90}$Zr at (a) $E=65$~MeV, (b) 185~MeV, (c) $E=400$~MeV and
(d) 800~MeV.
The solid, dashed and dotted curves stand for the
results of the BR local potential calculated with
choices (i), (ii) and (iii) for ${r}_g$, respectively.
See the text for details.
 }
 \label{fig:elastic-Zr-rg}
\end{center}
\end{figure}

Finally, we comment on the ${r}_g$-ambiguity of the BR-type local potential.
Figure \ref{fig:elastic-Zr-rg} shows the elastic-scattering cross section
from $^{90}$Zr at
(a) $E=65$~MeV, (b) 185~MeV, (c) $E=400$~MeV and (d) 800~MeV.
The solid lines represent results of the BR local potential
calculated with choice (i), and the dashed and dotted
lines correspond to results of
choices (ii) and (iii), respectively.
The differences among the three cases are appreciable
only at dips for $E=65$~MeV. For $E=185$~MeV, they are 
appreciable even for tops.
For higher $E$ such as $E \ga 400$~MeV, the differences become
negligible, since so is the medium effect itself.
Thus, the folding potential has an appreciable ${r}_g$-ambiguity
only around $E=200$~MeV in which the imaginary part of $U_{\rm BR}(R)$
is rather weak compared with the case of other $E$.

%%------------------------------------------------------------------%%
%% Figure
%%------------------------------------------------------------------%%
%%------------------------------------------------------------------%%

\section{Summary}
\label{Summary}

We test the BR localization of the microscopic nucleon-nucleus
optical potential
over the wide range of
$65 < E < 800$~MeV and conclude that the localization is valid there.
The BR localization is composed of
the local semi-classical approximation (LSCA),
the local Fermi-gas approximation (LFGA) and
Eq.~(\ref{SSA}),
but these approximations can be reduced to
two, LSCA and LFGA-$R$ (a modified version of LFGA).
The former is reliable at $E \ga 65$~MeV, while the latter is good for any
$E$.
The approximate wave functions calculated with
the BR-type local potential are very close to the exact ones.
Thus, the BR-type local potential is
quite useful in many applications, for example, as potentials
between A and constituents of weakly bound or unstable projectiles.

\section*{Acknowledgements}
The authors thank Kawai, Sakuragi and Nakada
for their interest to the present
subject.

%%%%%%%%%%%%%%%%%%%%%%%

%%%%%%%%%%%%%%%%  References %%%%%%%%%%%%%%%%%%

\end{document}